\documentstyle[12pt]{article}

\begin{document}

\setcounter{page}{1}

\pagestyle{plain} \vspace{1cm}
\begin{center}
\Large{\bf Crossing the Cosmological Constant Line on the Warped DGP Brane}\\
\small \vspace{1cm} {\bf Kourosh
Nozari$^{a,b,}$\footnote{knozari@umz.ac.ir}}, \quad {\bf Noushin
Behrouz$^{c,}$\footnote{noushin.behrouz@gmail.com}},\quad {\bf
Tahereh Azizi$^{a,}$\footnote{t.azizi@umz.ac.ir}}\\ and\\
\quad {\bf Behnaz Fazlpour$^{a,}$\footnote{b.fazlpour@umz.ac.ir}} \\
\vspace{0.5cm} {\it $^{a}$Department of Physics, Faculty of Basic
Sciences,\\
University of Mazandaran,\\
P. O. Box 47416-95447, Babolsar, IRAN\\
$^{b}$Research Institute for Astronomy and Astrophysics of Maragha,
\\P.
O. Box 55134-441, Maragha, IRAN }\\
{\it $^{c}$ Department of Physics, Payam-e Nour University,\\
P. O. Box 919, Mashad, IRAN}
\end{center}
\vspace{1.5cm}
\begin{abstract}
We study dynamics of the equation of state parameter for a dark
energy component non-minimally coupled to induced gravity on a
warped DGP brane. We show that there are appropriate domains of the
model parameters space that account for crossing of the phantom
divide line. This crossing, which is possible for both branches of
the scenario, depends explicitly on the values of the non-minimal
coupling and warp factor. The effect of warp factor appears in the
value of the redshift parameter at which phantom divide line crossing occurs. 
\end{abstract}
\vspace{2cm}
\newpage
\section{Introduction}
Recent observational data from CMB temperature fluctuations
spectrum, Supernova type Ia redshift-distance surveys and other data
sources, have shown that the universe is currently in a positively
accelerated phase of expansion and its spatial geometry is nearly
flat\, [1]. Nevertheless, there is not enough standard matter
density in the universe to support this flatness and accelerated
expansion. Therefore, we need either additional cosmological
components or modify general relativity at cosmological scales to
explain these achievements [2]. Multi-component dark energy with at
least one non-canonical phantom field\footnote{ It is important to
note that phantom fields are not consistent due to violation of the
null energy condition and instabilities ( see for instance [3]).
However, theoretically they provide a good candidate with negative
pressure to realize late-time accelerated expansion. Recently, it
has been shown that phantom-like behavior can be realized without
introducing any phantom matter in some specific braneworld models
[4]. We note also that in our model due to its wider parameter
space, it is expected essentially that the null energy condition is
fulfill in at least some subspaces of the model parameter space. }
is a possible candidate of the first alternative. This viewpoint has
been studied extensively in literature ( see [5,6] and references
therein ). There are some datasets (such as the Gold dataset) that
show a mild trend for crossing of the phantom divide line by
equation of state (EoS) parameter of dark component. The equation of
state parameter in these scenarios crosses the phantom divide line (
$\omega=\frac{p}{\rho}=-1$) at recent redshifts and current
accelerated expansion requires $\omega<-\frac{1}{3}$.  In fact,
recent observational data restrict $\omega(z)$ to be larger than
$-1$ in the past and less than $-1$ today. The current best fit
value of the equation of state parameter, using WMAP five year data
combined with measurements of Type Ia supernovae and Baryon Acoustic
Oscillations in the galaxy distribution, is given by $-0.11 <
1+\omega < 0.14$ ( with $95$ percent CL uncertainties) [7]. It is
accepted that crossing of the phantom divide line occurs at recent
epoch with $z\sim0.25$ [5,6], although this value is model
dependent. Currently, models of phantom divide line crossing are so
important that they can realize that which model is better than the
others to describe the nature of dark energy [6]. Although this
crossing cannot be explained just by one scalar field [8],
generalization to multi-field case or non-minimal coupling with
gravity provide enough space to achieve such a crossing [3,5,6].
Lorentz invariance violating fields are other alternative dark
energy components with capability to cross the phantom divide line
by EoS parameter in a fascinating manner [9]. Also it is possible to
reconstruct a scalar-tensor theory of gravity in an accelerating
universe where a phantom behavior can realized [10]. The
cosmological constant ( with $\omega(z)=-1$) is the simplest
candidate for dark energy [11,12]. However this scenario suffers
from some difficulties such as lack of physical motivation, huge
amount of fine-tuning to explain cosmological accelerated expansion
and no dynamics for its equation of state [12]. So it seems
worthwhile to probe alternative dynamical models.

Another alternative to explain current accelerated expansion of the
universe is extension of the general relativity to more general
theories on cosmological scales ( see [13] and references therein).
DGP ( Dvali-Gabadadze-Porrati) braneworld scenario as an infra-red
(IR) modification of general relativity, explains accelerated
expansion of the universe in its self-accelerating branch via
leakage of gravity to extra dimension [14]. In this model, the
late-time acceleration of the universe is driven by the
manifestation of the excruciatingly slow leakage of gravity off our
four-dimensional world into an extra dimension [15]. In this
scenario the EoS parameter of dark energy never crosses the
$\omega(z)=-1$ line, and universe eventually turns out to be de
Sitter phase. However, in this setup if we use a single scalar field
(ordinary or phantom) on the brane, we can show that EoS parameter
of dark energy can cross the phantom divide line [16]. It has been
shown that DGP model with a quintom dark energy fluid ( a
combination of quintessence and phantom fields in a joint model) in
the bulk or brane, accounts for accelerated expansion and the
phantom divide line crossing [17]. In a braneworld setup with
induced gravity embedded in a bulk with arbitrary matter content,
the transition from a period of domination of the matter energy
density by nonrelativistic brane matter to domination by the
generalized dark radiation provides a crossing of the phantom divide
line [18]. In this setup there is no need to introduce additional
scalar field as a dark component. Recently, phantom-like behavior in
a brane-world setup with induced gravity and also curvature effects
have been reported [19]. On the other hand, Gauss-Bonnent braneworld
scenario with induced gravity dose not need introducing any scalar
field to account for this crossing. In other words, the combination
of the effect of Gauss-Bonnent term in the bulk and induced gravity
term on the brane behaves as dark energy on the brane [20].
Quintessential scheme can also be achieved in a geometrical way in
higher order theories of gravity [21].

Crossing of the phantom divide line by a minimally coupled scalar
field on the DGP braneworld has been studied by Zhang and Zhu [16].
In this setup, there are two possible cases: for ordinary scalar
field EoS of dark energy crosses from $\omega>-1$ to $\omega<-1$ in
normal ( non self-accelerating) branch and for phantom field EoS of
dark energy crosses from $\omega<-1$ to $\omega>-1$ in
self-accelerating branch of DGP scenario. As an important
generalization, the Randall-Sundram II model [22] combined with DGP
scenario provides a rich structure which has been called {\it
warped} DGP braneworld in literature [23]. In this model, an induced
curvature term appears on the brane in the RS II model.

Our motivation in this paper is to consider a scalar field
non-minimally coupled to induced gravity on the warped DGP
braneworld as a dark energy component and investigate the roles
played by non-minimal coupling and the warp effect in the dynamics
of the equation of state parameter. We study dynamics of the
equation of state parameter focusing on the crossing of the phantom
divide line in this setup. We show that this crossing is possible
for a suitable range of the model parameters and especially for some
specific values of the non-minimal coupling and warp factor. More
specifically, we show that for ordinary scalar (quintessence) field
in the self-accelerating branch of the warped DGP model and with
positive values of the non-minimal coupling, the EoS of dark energy
runs from below $-1$ ($\omega_{de}<-1$) to above $-1$. In normal
branch of this warped DGP model with positive and negative values of
the non-minimal coupling, the EoS parameter runs from above $-1$ to
below $-1$ which is supported by recent observations. For phantom
field we have crossing of the phantom divide line in both branches
of the warped DGP setup and in both of these cases the EoS parameter
runs from above $-1$ to below $-1$. But, in self-accelerating branch
of the model we have crossing behavior with positive and negative
values of the non-minimal coupling in normal branch and crossing
occurs just with negative values of the non-minimal coupling.

\section{A Dark Energy Model on the Warped DGP Brane}
\subsection{Warped DGP Brane}
Let us start with the action of the warped DGP model as follows
\begin{equation}
{\cal{S}}={\cal{S}}_{bulk}+{\cal{S}}_{brane},
\end{equation}
\begin{equation}
{\cal{S}}=\int_{bulk}d^{5}X\sqrt{-{}^{(5)}g}\bigg[\frac{1}{2\kappa_{5}^{2}}
{}^{(5)}R+{}^{(5)}{\cal{L}}_{m}\bigg]+\int_{brane}d^{4}x\sqrt{-g}\bigg[\frac{1}{\kappa_{5}^{2}}
K^{\pm}+{\cal{L}}_{brane}(g_{\alpha\beta},\psi)\bigg].
\end{equation}
Here ${\cal{S}}_{bulk}$ is the action of the bulk,
${\cal{S}}_{brane}$ is the action of the brane and ${\cal{S}}$ is
the total action. $X^{A}$ with $A=0,1,2,3,5$ are coordinates in the
bulk, while $x^{\mu}$ with $\mu=0,1,2,3$ are induced coordinates on
the brane. $\kappa_{5}^{2}$ is the $5$-dimensional gravitational
constant. ${}^{(5)}R$ and ${}^{(5)}{\cal{L}}_{m}$ are
$5$-dimensional Ricci scalar and matter Lagrangian respectively.
$K^{\pm}$ is trace of the extrinsic curvature on either side of the
brane. ${\cal{L}}_{brane}(g_{\alpha\beta},\psi)$  is the effective
4-dimensional Lagrangian. The action ${\cal{S}}$ is actually a
combination of the Randall-Sundrum II and DGP model. In other words,
an induced curvature term is appeared on the brane in the
Randall-Sundrum II model, hence the name {\it warped} DGP Braneworld
[23]. Now consider the brane Lagrangian as follows
\begin{equation}
{\cal{L}}_{brane}(g_{\alpha\beta},\psi)=\frac{\mu^2}{2}R-\lambda+L_{m},
\end{equation}
where $\mu$ is a mass parameter, $R$ is Ricci scalar of the brane,
$\lambda$ is tension of the brane and $L_{m}$ is Lagrangian of the
other matters localized on the brane. Assume that bulk contains only
a cosmological constant, $^{(5)}\Lambda$. With these choices, action
(1) gives either a generalized DGP or a generalized RS II model: it
gives DGP model if $\lambda=0$ and $^{(5)}\Lambda=0$, and gives RS
II model if $\mu=0$ [23]. The generalized Friedmann equation on the
brane is as follows [23]
\begin{equation}
H^{2}+\frac{k}{a^{2}}=\frac{1}{3\mu^2}\bigg[\rho+\rho_{0}\Big(1+\varepsilon
{\cal{A}}(\rho,a)\Big)\bigg],
\end{equation}
where $\varepsilon=\pm 1$ is corresponding to two possible branches
of the solutions in this warped DGP model and
${\cal{A}}=\bigg[{\cal{A}}_{0}^{2}+\frac{2\eta}{\rho_{0}}
\Big(\rho-\mu^{2}\frac{{\cal{E}}_{0}}{a^{4}}\Big)\bigg]^{1/2}$ where
\,\, ${\cal{A}}_{0}\equiv
\bigg[1-2\eta\frac{\mu^{2}\Lambda}{\rho_{0}}\bigg]^{1/2}$,\,\, $\eta
\equiv\frac{6m_{5}^{6}}{\rho_{0}\mu^{2}}$\,\, with $0<\eta\leq1$
which is related to the warped geometry of the bulk manifold \,\,and
\,\,$\rho_{0}\equiv m_{\lambda}^{4}+6\frac{m_{5}^{6}}{\mu^{2}}$. By
definition, $m_{\lambda}= \lambda^{1/4}$ and $m_{5}=k_{5}^{-2/3}$.\,
${\cal{E}}_{0}$ is an integration constant and corresponding term in
the generalized Friedmann equation is called dark radiation term. We
neglect dark radiation term in which follows. In this case,
generalized Friedmann equation (4) attains the following form,
\begin{equation}
H^{2}+\frac{k}{a^2}=\frac{1}{3\mu^2}\bigg[\rho+\rho_{0}+\varepsilon
\rho_{0}\Big({\cal{A}}_{0}^{2}+\frac{2\eta\rho}{\rho_{0}}\Big)^{1/2}\bigg],
\end{equation}
where $\rho$ in our case is the total energy density, including
scalar field and dust matter on the brane
\begin{equation}
\rho=\rho_\varphi+\rho_{dm}.
\end{equation}
In which follows we construct a dark energy model on the warped DGP
setup.

\subsection{ Quintessence Field}

Now we consider a quintessence scalar field non-minimally coupled to
induced gravity on the warped DGP brane as a candidate for dark
energy. The action of this non-minimally coupled scalar field is
given by
\begin{equation}
{\cal{S}}_{\varphi}=\int_{brane}d^{4}x\sqrt{-g}\Big[-\frac{1}{2}\xi
R\varphi^{2}-\frac{1}{2}\partial_{\mu}\varphi\partial^{\mu}\varphi-V(\varphi)\Big],
\end{equation}
where $\xi $ is a non-minimal coupling and $R$ is Ricci scalar of
the brane. We have assumed a conformal coupling of the scalar field
and induced gravity. The scalar field will play the role of
dark-energy component on the brane.  Variation of the action with
respect to $\varphi$ gives the equation of motion of the scalar
field
\begin{equation}
\ddot{\varphi}+3H\dot{\varphi}+\xi R\varphi +\frac{dV}{d\varphi}=0.
\end{equation}
The energy density and pressure of the non-minimally coupled scalar
field are given by
\begin{equation}
\rho_{\varphi}=\frac{1}{2}\dot{\varphi}^{2}+V(\varphi)+6\xi
H\varphi\dot{\varphi}+3\xi H^{2}\varphi^{2}
\end{equation}
\begin{equation}
p_{\varphi}=\frac{1}{2}\dot{\varphi}^{2}-V(\varphi)-2\xi(\varphi\ddot{\varphi}+2\varphi
H\dot{\varphi}+\dot{\varphi}^{2})-\xi\varphi^2(2\dot{H}+3H^2)
\end{equation}
In which follows, by comparing the modified Friedmann equation in
the warped DGP braneworld with the standard Friedmann equation, we
deduce a definition for equation of state of dark energy component.
This is reasonable since all observed features of dark energy are
essentially derivable in general relativity [16,20]. The standard
Friedmann equation in four dimensions is written as
\begin{equation}
H^2+\frac{k}{a^2}=\frac{1}{3\mu^2}(\rho_{dm}+\rho_{de}),
\end{equation}
where $\rho_{dm}$ is the dust matter density, while $\rho_{de}$ is
dark energy density. Comparing this equation with equation (5) for a
spatially flat universe ( $k=0$), we find
\begin{equation}
\rho_{de}=\rho_{\varphi}+\rho_0+\varepsilon\rho_0\Big(A_0^2+2\eta\frac{\rho}{\rho_0}\Big)^{\frac{1}{2}}.
\end{equation}
Conservation of the scalar field effective energy density leads to
\begin{equation}
\frac{d\rho_{\varphi}}{dt}+3H(\rho_{\varphi}+p_{\varphi})=0.
\end{equation}
We note that the non-minimal coupling of the scalar field to the
Ricci curvature on the brane preserves conservation of the scalar
field energy density\footnote{Note that authors of Ref. [24] have
treated this conservation in relatively different way. They have
defined a total energy-momentum tensor consist of two parts: a pure
(canonical) scalar field energy-momentum tensor and a non-minimal
coupling-dependent part. The total energy density defined in this
manner is then conserved. In our case, we have included all possible
terms in equations (9) and (10) from beginning and evidently total
energy density defined in this manner is conserved too. In fact, it
is simple to show that our equations (9) and (10) are equivalent to
$\rho^{tot}$ and $P^{tot}$ of Ref. [24] if we set
$\alpha(\varphi)=\frac{1}{2}(1-\xi\varphi^{2})$ ( see also [25] for
more detailed discussion).}.

Since the dust matter obeys the continuity equation and the Bianchi
identity keeps valid, dark energy itself satisfies the continuity
equation
\begin{equation}
\frac{d\rho_{de}}{dt}+3H(\rho_{de}+p_{de})=0
\end{equation}
where $p_{de}$ denotes the pressure of the dark energy. Note that
this is just a definition and by assuming validity of this
definition we can obtain effective pressure of dark energy as well
as an effective equation of state. Now, the equation of state for
this dark energy component can be written as follows
\begin{equation}
w_{de}=\frac{p_{de}}{\rho_{de}}=-1+\frac{1}{3}\frac{d\ln\rho_{de}}{d\ln(1+z)}\,,
\end{equation}
where by using (13) and (14) we find
\begin{equation}
\frac{d\ln\rho_{de}}{d\ln(1+z)}=\frac{3}{\rho_{de}}
\bigg[\rho_{\varphi}+p_{\varphi}+\varepsilon\eta\bigg(A_0^2+
2\eta\frac{\rho_{\varphi}+\rho_{dm}}{\rho_0}\bigg)^{-\frac{1}{2}}\bigg(\rho_{\varphi}
+p_{\varphi}+\rho_{dm}\bigg)\bigg].
\end{equation}
There are three possible alternatives in this setup: if
$\frac{1}{3}\frac{d\ln\rho_{de}}{d\ln(1+z)}>0$, we have a
quintessence model; if
$\frac{1}{3}\frac{d\ln\rho_{de}}{d\ln(1+z)}<0$, the model is phantom
and if $\frac{1}{3}\frac{d\ln\rho_{de}}{d\ln(1+z)}=0$, the dark
component is a cosmological constant. Evidently, in this setup
non-minimal coupling of the scalar field and induced gravity plays a
crucial role supporting or preventing phantom divide line crossing.
In this respect, the differences between the minimal and non-minimal
setups will be more clear if we write the explicit dynamics of the
equation of state parameter. We also discuss the effect of warp
factor on the dynamics of the equation of state parameter in
forthcoming arguments. We choose the following exponential potential
with motivation that this type of potential can be solved exactly in
the standard model
\begin{equation}
 V=V_0 \exp(-\lambda\frac{\varphi}{\mu}),
\end{equation}
where $V_0$, $\lambda$ and $\mu$ are constant.\\
Differentiation of the logarithm of dark energy effective density
with respect to $ln(1+z)$ yields
$$\frac{d\ln\rho_{de}}{d\ln{(1+z)}}=\frac{3}{\rho_{de}}\Bigg[\dot{\varphi}^2-2\xi\Big(-H\varphi\dot{\varphi}
+\dot{H}\varphi^2+\varphi\ddot{\varphi}+\dot{\varphi}^2\Big)
+\Big[\dot{\varphi}^2-2\xi\Big(-H\varphi\dot{\varphi}
+\dot{H}\varphi^2+\varphi\ddot{\varphi}+\dot{\varphi}^2\Big)+\rho_{dm}\Big]$$
\begin{equation}
\Big[\varepsilon\eta\Big(A_{0}^2+2\eta\frac{\frac{1}{2}\dot{\varphi}^{2}+V(\varphi)+6\xi
H\varphi\dot{\varphi}+3\xi
H^{2}\varphi^{2}+\rho_{dm}}{\rho_0}\Big)^{-\frac{1}{2}}\Big]\Bigg].
\end{equation}
To study the behavior of the EoS parameter of dark energy component,
we need to the explicit form of $\ddot{\varphi}$ in terms of other
quantities which can be deduced from equation of motion as given by
(8). On the other hand, Friedmann equation given by (5) now takes
the following form
$$ (\mu^2+g)^{2}H^4+2f(3\mu^2+g)H^3+\Big[f^2-2l(3\mu^2+g)+2\eta\rho_0 g\Big]H^2$$
\begin{equation}
+\Big(-2fl+\rho_0\eta f\Big)H-2\eta\rho_0(l-\rho_0)-\rho_0^2
A_{0}^2+l^2=0
\end{equation}
where
$$g=-3\xi H^{2}\varphi^{2},$$
$$l=\frac{1}{2} \dot{\varphi^2}+V(\varphi)+\rho_{dm}+\rho_{0},$$
and
$$f=-6\xi H\varphi\dot{\varphi}$$
Equation (20) is a quadratic equation in terms of $H^{2}$ and in
principle has four roots for $H$. We show these roots as $h_{1}$,
$h_{2}$, $h_{3}$ and $h_{4}$. After numerical calculation, we found
that two of these roots, say, $h_{1}$ and $h_{2}$ are unphysical and
excluded by observational data. The other two roots, $h_{3}$ and
$h_{4}$, are physical solutions corresponding to the generalized
normal branch (with $\varepsilon=-1$) and the self-accelerating one
(with $\varepsilon=-1$). As we will show these solutions have the
capability to account for phantom divide line crossing. We introduce
a new parameter defined as $s=-\ln (1+z)$ and rewrite dark energy
equation of state parameter as follows
\begin{equation}
w_{de}=-1-\frac{1}{3}\frac{d\ln{\rho_{de}}}{d\ln{s}}.
\end{equation}
Now, we analyze the behavior of $w_{de}$ versus $s$ to investigate
cosmological implications of this scenario. Using equation (18), we
see that in the minimal case ( with $\xi=0$ ) and neglecting warp
effect, if we choose the sign of $\varepsilon$ to be negative,
remaining terms have suitable combination of signs so that it is
possible to cross the phantom divide line by the EoS parameter. In
the non-minimal case, however, it is not simple to conclude that
there is crossing of the phantom divide line or not just by defining
the sign of $\varepsilon$, since in this case non-minimal coupling
itself has a crucial role in the dynamics of the equation of state
parameter. We consider $\xi$ as a fine-tuning parameter in this
case. It is important to note that in the absence of the scalar
field on the DGP setup, there is no crossing of the phantom divide
line on the self-accelerating branch even in the warped DGP
scenario. Nevertheless, normal branch accounts for crossing of the
phantom divide line in this situation. In which follows, to
calculate EoS parameter $\omega(z)$, we define some dimensionless
density parameters such as ( see the paper by Sahni and Shtanov in
Ref. [4] and also Ref. [15])
\begin{equation}
\Omega_{\alpha}\equiv\frac{\rho_{0}^{(\alpha)}}{3\mu^{2}H_{0}^{2}},
\end{equation}
where we have assumed that $\rho$ is given by the sum of the energy
densities $\rho_{\alpha}$ of different components labeled by
$\alpha$ with a constant EoS parameter, $\omega_{\alpha}$. Also we
define
\begin{equation}
\Omega_{k}\equiv-\frac{k}{H_{0}^{2}a_{0}^{2}},\quad\quad
\Omega_{r_{c}}\equiv \frac{1}{H_{0}^{2}r_{c}^{2}},\quad\quad
\Omega_{\lambda}\equiv\frac{\lambda}{3\mu^{2}H_{0}^{2}}, \quad\quad
\Omega_{\Lambda}\equiv-\frac{^{(5)}\Lambda}{6H_{0}^{2}},
\end{equation}
where $r_{c}$ is DGP crossover distance. In this case, Friedmann
equation can be written as follows

$$\frac{H^{2}(z)}{H_{0}^{2}}=\Omega_{k}(1+z)^{2}+\Omega_{dm}
\big[(1+z)^{3}-1\big]+\Omega_{\phi}\big[(1+z)^{3(1+\omega)}-1\big]+\Omega_{\lambda}+2\Omega_{r_{c}}$$
\begin{equation}
\pm2\bigg[\Omega_{r_{c}}\bigg(\Omega_{dm}\big[(1+z)^{3}-1\big]+
\Omega_{\phi}\big[(1+z)^{3(1+\omega)}-1\big] +
\Omega_{\lambda}+\Omega_{r_{c}}+\Omega_{\Lambda}\bigg)\bigg]^{1/2}
\end{equation}
where $\pm$ stands for two possible embedding of the brane in the
bulk. The constraint equations for cosmological parameters are given
by
\begin{equation}
1-\Omega_{k}+\Omega_{\Lambda}=
\bigg[\sqrt{\Omega_{r_{c}}+\Omega_{dm}+\Omega_{\phi}+\Omega_{\lambda}
+\Omega_{\Lambda}}\pm\sqrt{\Omega_{r_{c}}}\bigg]^{2}.
\end{equation}
We can define also $\Omega_{ki}$ as the present value of the scalar
field kinetic energy density $\frac{1}{2}\dot{\varphi}^{2}$ over the
critical density defined as $\rho_{c}=3H_{0}^{2}$ ( with $8\pi
G=1$). Note that $\Omega_{ki}$ and the non-minimal coupling
parameter are hidden in the definition of the $\Omega_{\phi}$. If we
change the values of these parameters in appropriate manner, the
redshift at which crossing of the phantom divide line occurs will
change since it is a model dependent quantity in this respect. In
table $1$, we have obtained some reliable ranges of the non-minimal
coupling to have crossing of the phantom divide line in this setup.
Observational data show that crossing of the phantom divide line is
occurred in redshift \footnote{ Note that this is a model dependent
value but this value is suitable for our purposes in forthcoming
arguments.} $z\simeq 0.25$, so we have obtained the values of $\xi$
which are correspond to this redshift in the last column of the
table $1$. We have not excluded the negative values of the
non-minimal coupling from our analysis. In fact, these negative
values are theoretically interesting, corresponding to
anti-gravitation ( see [25] for further discussion).
\begin{table}
\begin{center}
\caption{Acceptable range of $\xi$ to have crossing of the phantom
divide line with quintessence field ( constraint by the age of the
universe). } \vspace{0.5 cm}
\begin{tabular}{|c|c|c|c|c|c|c|c|}
  \hline
  \hline $\varepsilon$&$\xi$ &Acceptable range of $\xi$ & The value of $\xi$ for z=0.25  \\
  \hline +1&negative & $-0.605<\xi\leq0$ &-0.438 \\
  \hline +1&positive & no crossing & --- \\
 \hline -1&negative &$-0.83<\xi\leq0$ & -0.522 \\
  \hline-1&positive &$0\leq\xi<0.148$&0.124\\
 \hline
\end{tabular}
\end{center}
\end{table}
The results of the numerical calculations are shown in figures $1$,
$2$, $3$ and $4$ for two branches of this DGP-inspired model and
with different values of the non-minimal coupling $\xi$. In this
figures, the best ranges of the values for $\xi$ to have a reliable
model in comparison with observational data are obtained. Note that
in all of our numerical calculations we have assumed
$\Omega_{ki}=0.01$, $\Omega_{r_{c}}=0.01$, $\Omega_{m}=0.3$,
$\Omega_{k}=0$, $\Omega_{\lambda}=\Omega_{de}=0.7$,
$\Omega_{\Lambda}=1$,
$A_0=1$, $H_0=1$, $\mu=1$ and $\eta=0.99$.\\
Figure $1(a)$ shows that for self-accelerating branch of the model (
with $\varepsilon=+1$) and with negative values of the non-minimal
coupling, there is a crossing of the phantom divide line by the
equation of state parameter( note that self-accelerating branch of
this DGP-inspired model accounts for late-time accelerated
expansion). As we know, the EoS of a single scalar field in standard
relativity never crosses the phantom divide line ( see for instance
the paper by Vikman in Ref. [8]). Also, in DGP scenario with a
canonical scalar field on the brane, there is no crossing of the
phantom divide line in the self-accelerating branch of the model (
see [16] and [27]), but in our model we have such a crossing due to
the existence of the non-minimal coupling and warp effect. In our
case, if we choose for instance $\xi=- 0.438$, we have a phantom
divide line crossing at the point with $s=-0.22$ corresponding to
$z=0.25$ in agreement with observations. However, this crossing
occurs for negative values of the non-minimal coupling. On the other
hand, here the EoS runs from below $-1$ to above $-1$ ( from phantom
to quintessence phase) and therefore avoids big-rip singularity.

\begin{figure}[htp]
\begin{center}
\includegraphics{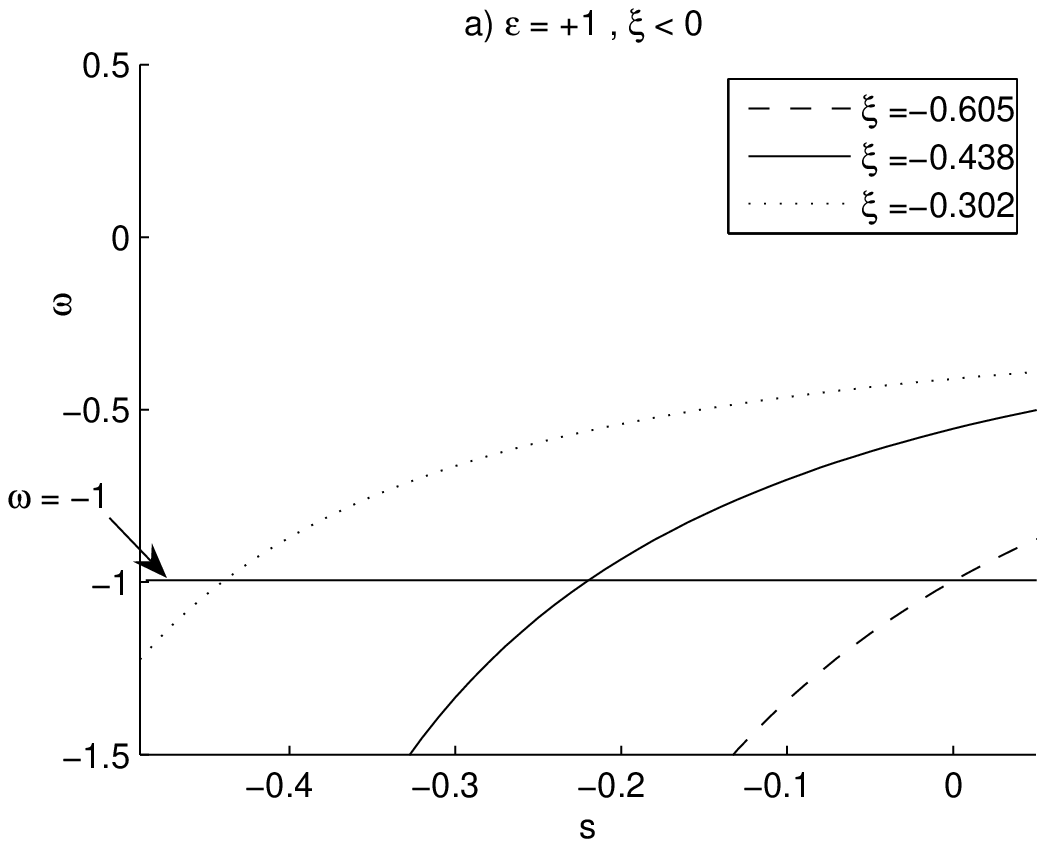} \vspace{5cm}\includegraphics{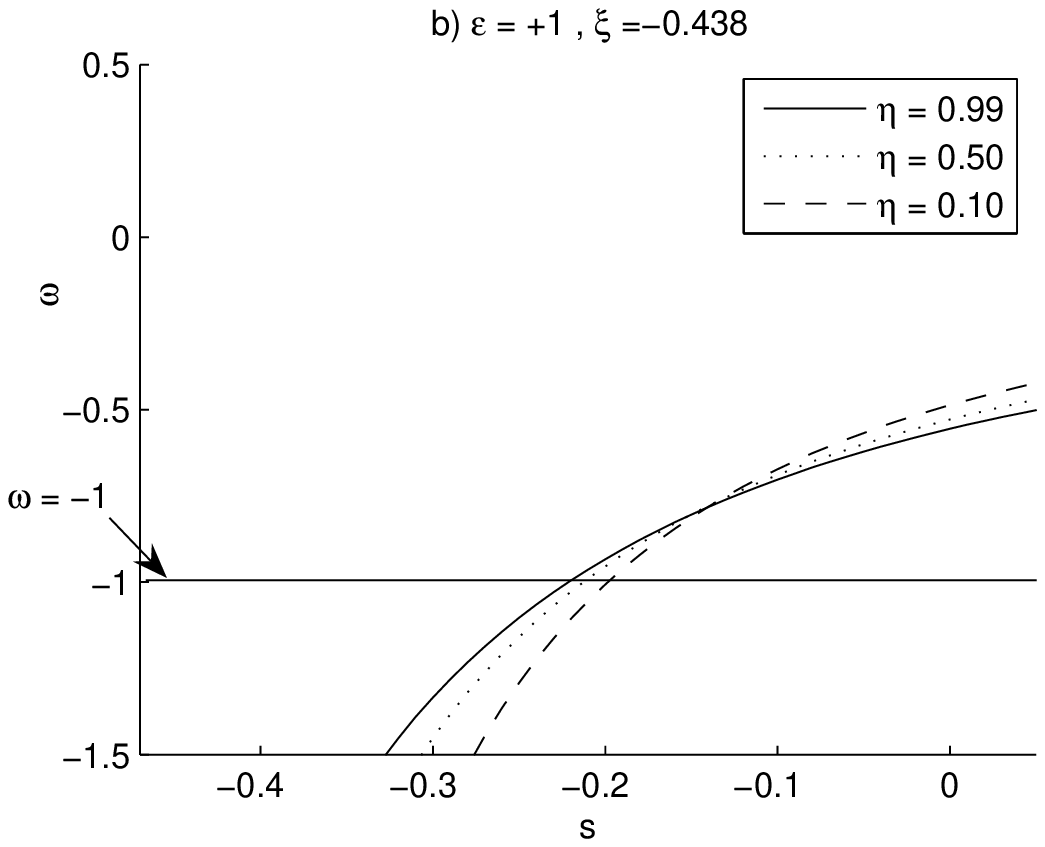}\vspace{5cm}\includegraphics{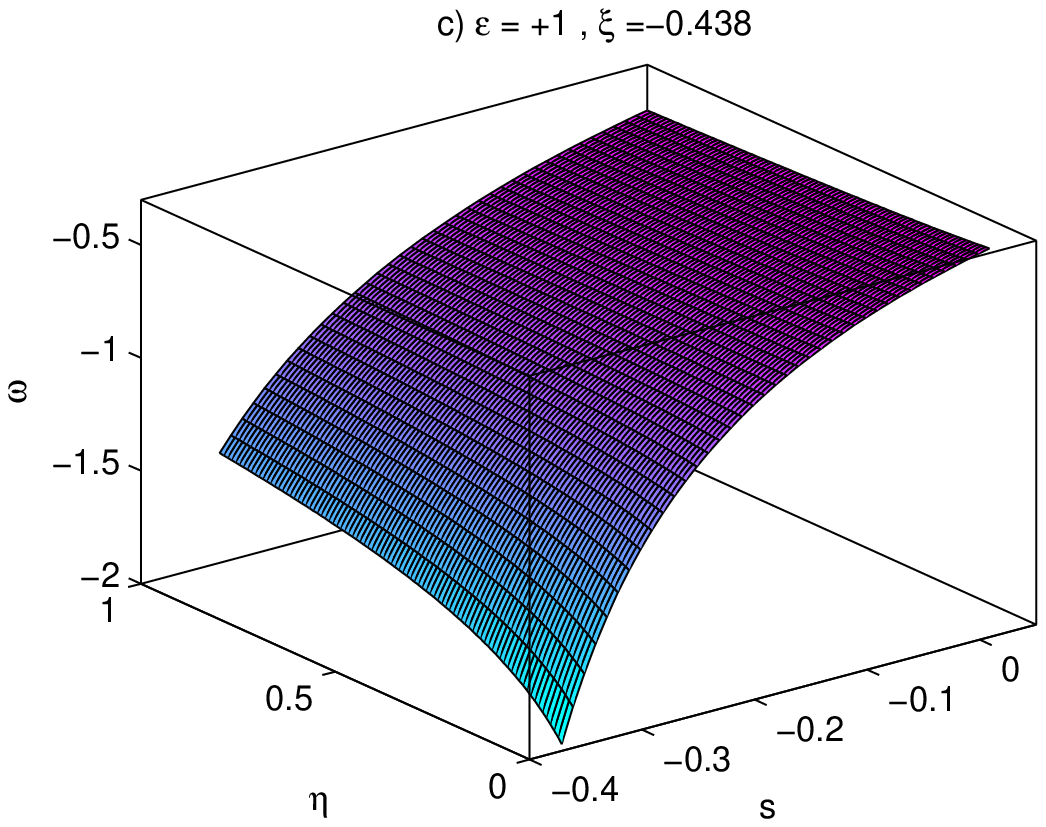}
\end{center}
\vspace{2.5cm}
 \caption{\small { a) In the self-accelerating branch of the model
 and with negative values of the nonminimal
 coupling, the EoS parameter crosses the phantom divide line. For instance,
 with $\xi=-0.438$ this crossing occurs at $s=-0.22$ or $z=0.25$.
 b)The role played by $\eta$ ( which is related to warp effect) on
 the crossing of the phantom divide line.
For sufficiently small values of $\eta$, equation of state
parameter, $\omega$, crosses the phantom divide line in relatively
small values of redshift. For example, with $\xi=-0.438$, the EoS of
dark energy crosses $\omega=-1$ line with $\eta=0.99$ at
$s\approx-0.22$ or $z=0.25 $, while for $\eta=0.50$ this crossing
occurs at $s\approx-0.212$ or $z=0.236$ and for $\eta=0.10$ this
crossing occurs at $s\approx-0.198$ or $z=0.218$. c) Equation of
state parameter, $\omega$, versus $s$ and $\eta$ with $\xi=-0.438$
in a three dimensional plot and within self-accelerating branch. In
self-accelerating branch with negative non-minimal coupling,
crossing of the phantom divide line occurs from phantom to
quintessence phase.}}
\end{figure}
The effect of the warp factor on the dynamics of the dark energy
component can be explained by the variation of $\eta$ parameter as
shown in figure $1(b)$. As this figure shows, for sufficiently small
values of $\eta$, equation of state parameter, $\omega$, crosses the
phantom divide line in relatively small values of redshift. In
figure $1(c)$ we plotted $\omega_{de}$ for $DGP^{(+)}$ branch of the
model with $\xi=-0.438$ with respect to the parameters $s$ and
$\eta$. In this figure, $\eta$ is restricted to the interval
$0<\eta\leq 1$.

\begin{figure}[htp]
\begin{center}
\includegraphics{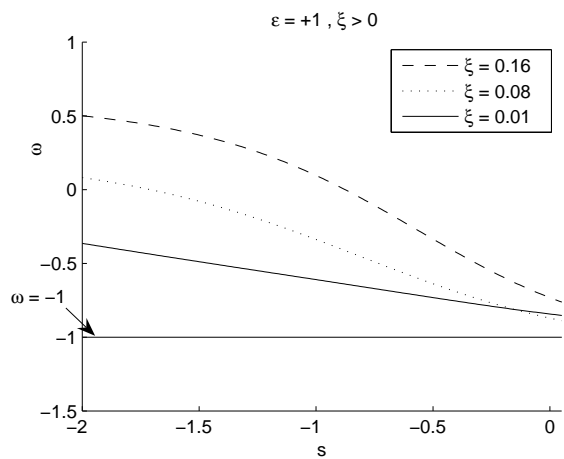}
\end{center}
\vspace{14 cm}
 \caption{\small {With a quintessence field on the warped DGP brane,
 there is no crossing of the phantom divide line in self-accelerating branch
 of the model with positive values of the non-minimal coupling.
 This means that inclusion of the warp factor cannot produce crossing of the
 phantom divide in this case. }}
\end{figure}
In figure $2$ we see that for self-accelerating branch of the model
and with positive values of the non-minimal coupling, there is no
crossing of the phantom divide line.

\begin{figure}[htp]
\begin{center}
\includegraphics{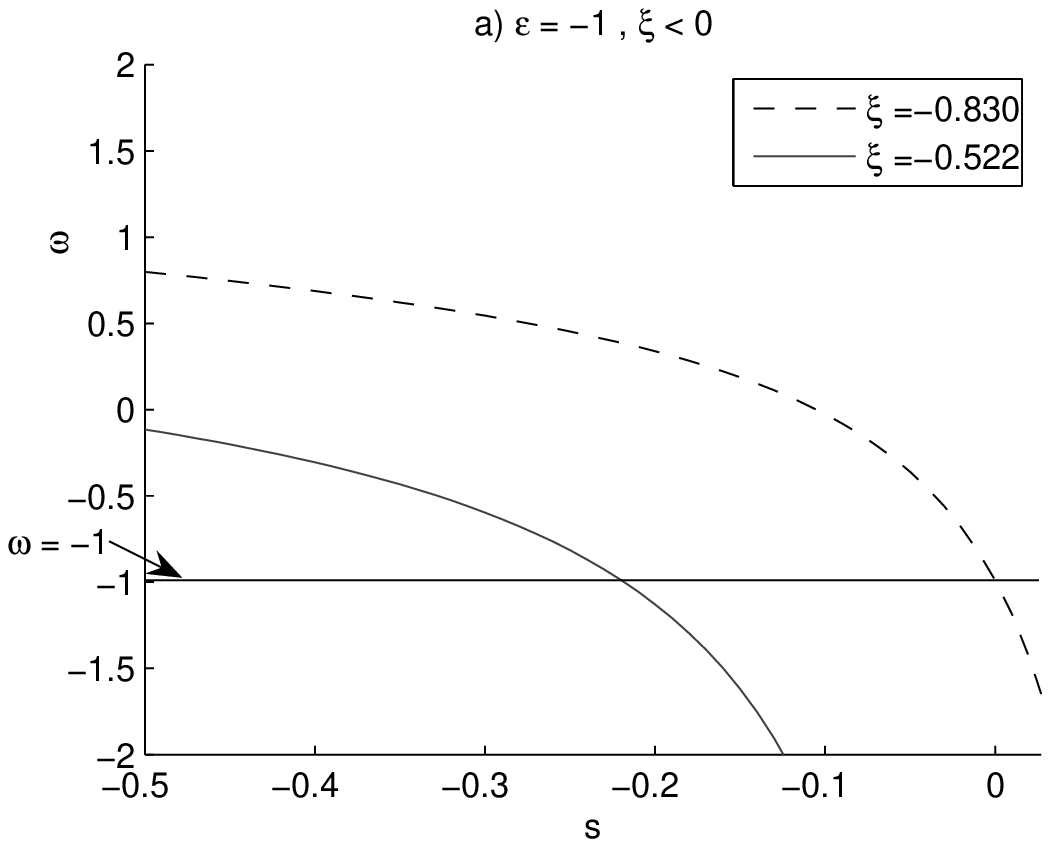} \vspace{5cm}\includegraphics{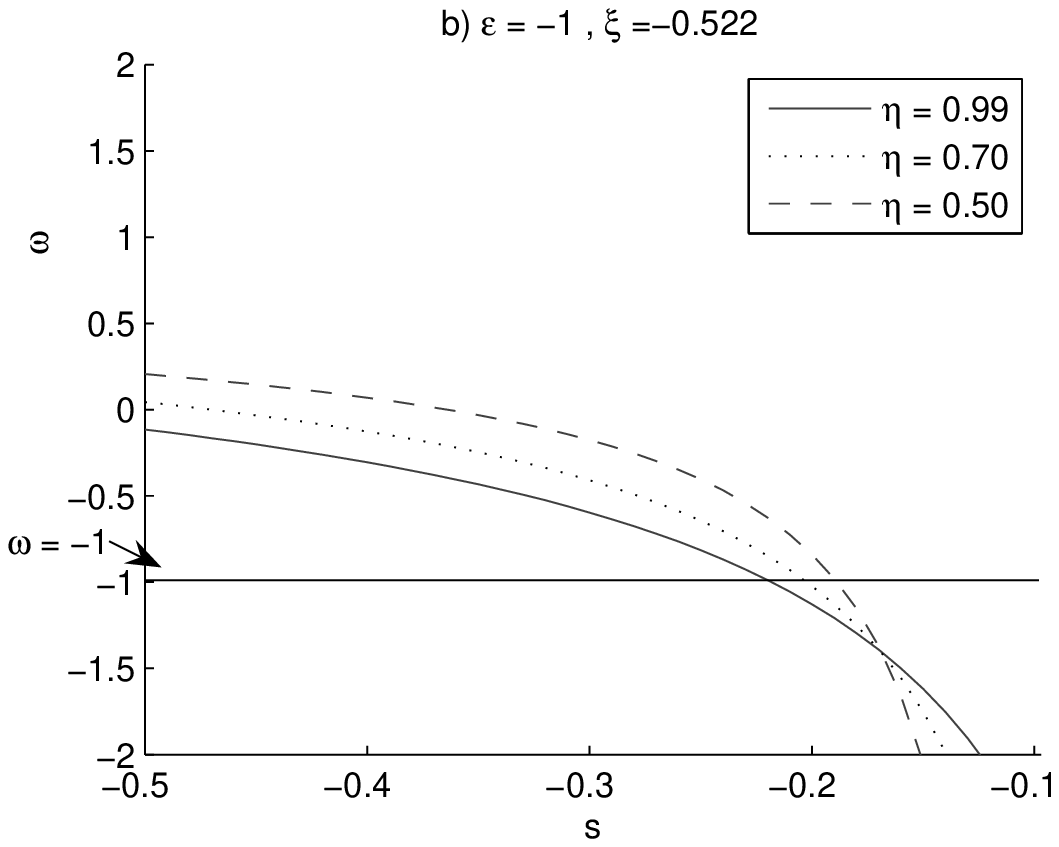}\vspace{5cm}\includegraphics{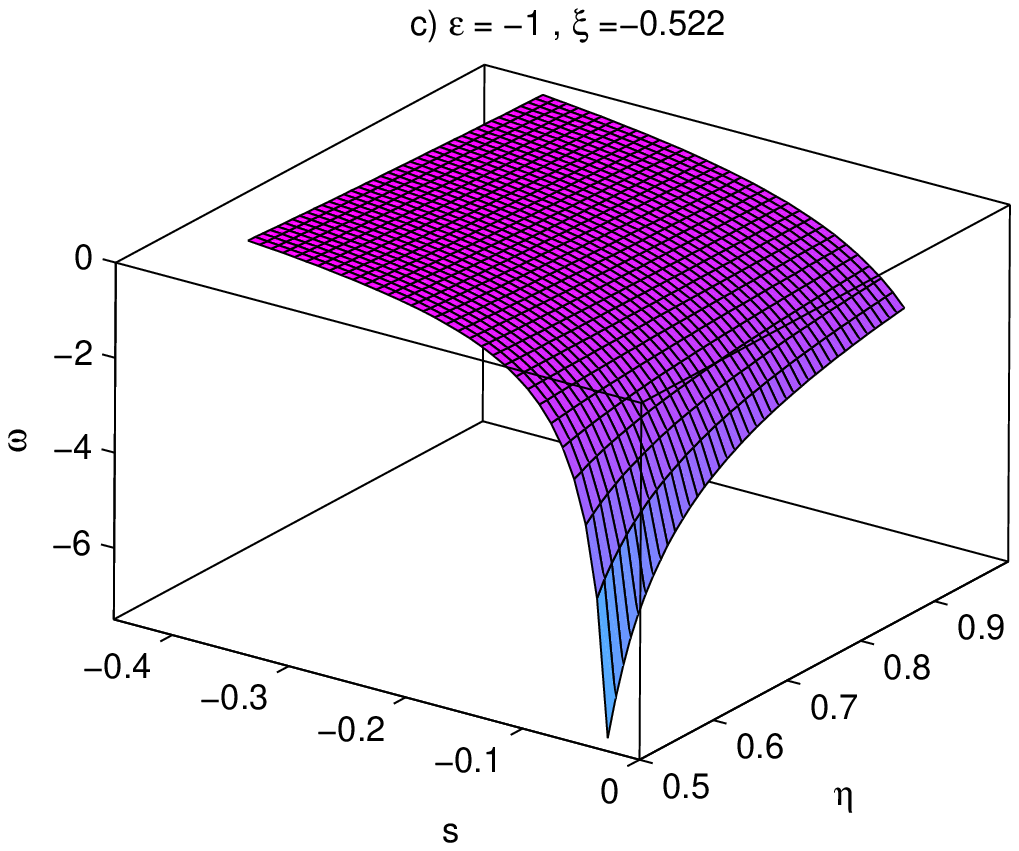}
\end{center}
\vspace{2.5cm}
 \caption{\small {a) In the normal branch of the model, there is
 a crossing of the phantom divide line with negative values of the nonminimal
 coupling. This crossing occurs from quintessence to phantom phase.
 For instance, the EoS parameter crosses the $\omega=-1$ line for
 $\xi=-0.522$ at $s=-0.22$ or $z=0.25$. b)The role played by $\eta$ factor.
 For sufficiently small values of $\eta$, equation of state
parameter, $\omega$, crosses the phantom divide line in relatively
small values of redshift. For example, with $\xi=-0.522$, the EoS of
dark energy crosses $\omega=-1$ line with $\eta=0.99$ at
$s\approx-0.22$ or $z=0.25$, while for $\eta=0.7$ this crossing
occurs at $s\approx-0.203$ or $z=0.225$ and for $\eta=0.5$ occurs at
$s\approx-0.189$ or $z=0.208$. c)Equation of state parameter
$\omega$, versus $s$ and $\eta$ with $\xi=-0.522$ in a
three-dimensional plot and for normal branch of the model. Note that
in this case, crossing runs from quintessence to phantom phase.}}
\end{figure}

Figure $3(a)$ shows that for normal branch ( with $\varepsilon=-1$)
of the model and with negative values of $\xi$, there is crossing of
the phantom divide line also. For instance, with $\xi=-0.522$, we
have a crossing at the point with $s=-0.22$ corresponding to
$z=0.25$ in agreement with observational data. The other note is
that EoS transits from $\omega_{de}>-1$ to $\omega_{de}<-1$ which is
supported by recent observation. Figure $3(b)$ shows that by
decreasing the values of $\eta$, equation of state parameter,
$\omega$, crosses the phantom divide line in relatively small values
of redshift. In figure $3(c)$ we plotted $\omega_{de}$ for
$DGP^{(-)}$ branch with
$\xi=-0.522$ and with respect to the parameters $s$ and $\eta$.\\
\begin{figure}[htp]
\begin{center}
\includegraphics{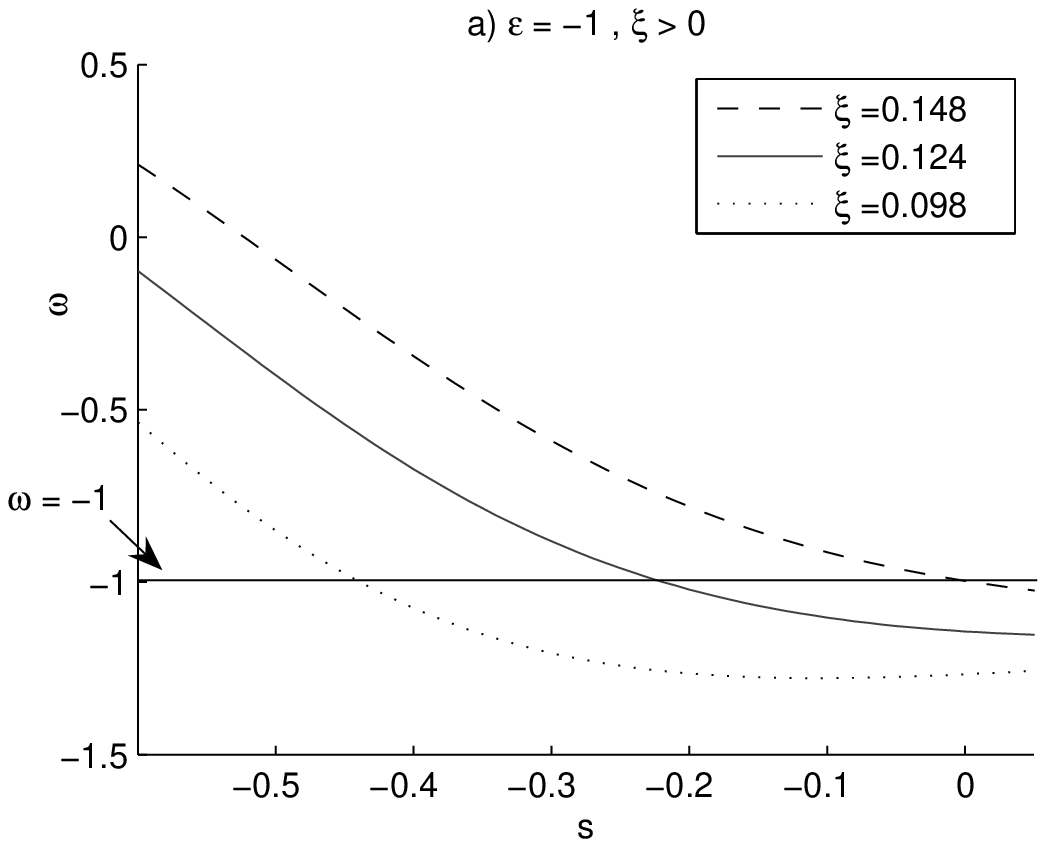} \vspace{5cm}\includegraphics{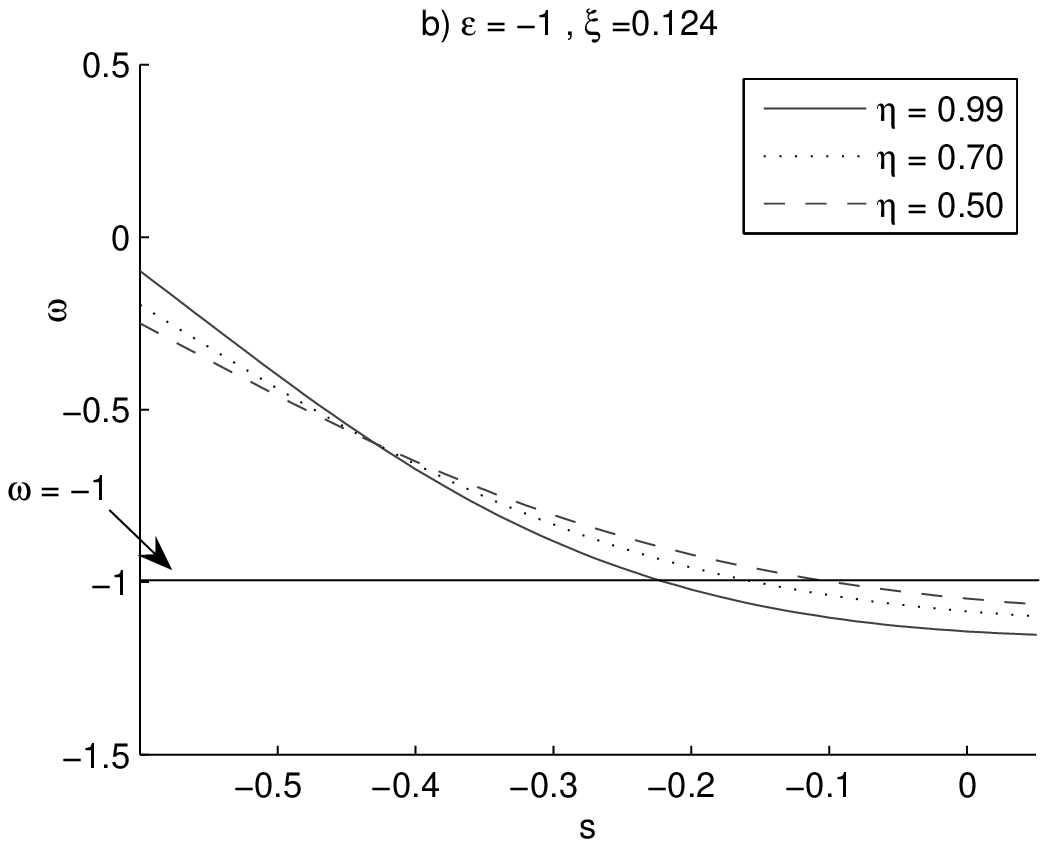}\vspace{5cm}\includegraphics{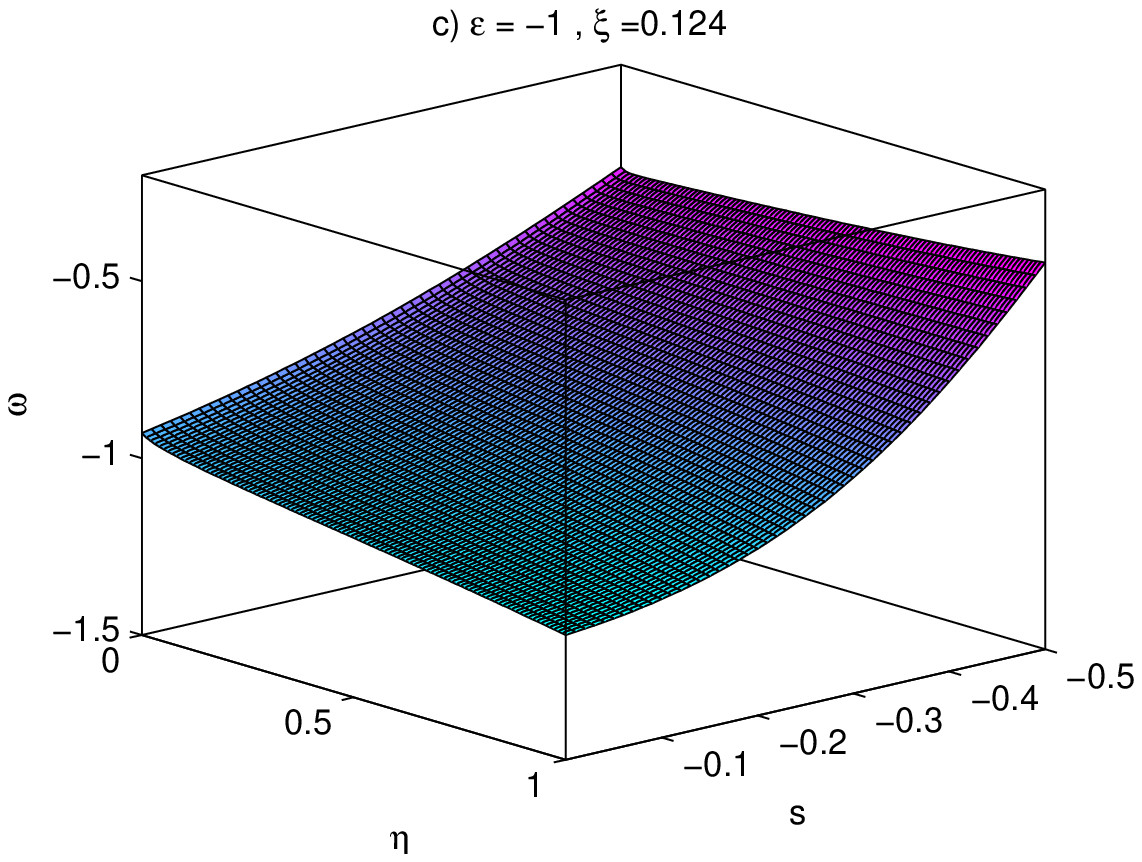}
\end{center}
\vspace{2.5cm}
 \caption{\small {The same as figure $3$ but now with positive values of the non-minimal coupling.
 a)In this case crossing of the phantom divide line runs from quintessence to phantom phase. For
 $\xi=0.124$ this crossing occurs at $s=-0.22$ or $z=0.25$. b)The role played by $\eta$ on the
 crossing of the phantom divide line. For sufficiently small values of $\eta$, equation of state
parameter crosses the phantom divide line in relatively small values
of redshift. For instance, if $\xi=0.124$, the EoS of dark energy
crosses $\omega=-1$ line for $\eta=0.99$ at $s\approx-0.22$, or
$z\approx0.25 $, while for $\eta=0.7$ this crossing occurs at
$s\approx-0.16$ or $z=0.173$ and for $\eta=0.5$ occurs at
$s\approx-0.10$ or $z\approx0.105$. c)Equation of state parameter,
$\omega$, versus $s$ and $\eta$ with $\xi=0.124$ in a 3-dimensional
plot. }}
\end{figure}
Figure $4(a)$ shows that EoS parameter of dark energy crosses the
phantom divide line in the normal branch ($\varepsilon=-1$) of the
model with positive $\xi$. For $\xi=0.124$ we have a crossing at the
point with $s=-0.22$ corresponding to $z=0.25$. Here, the EoS of
dark energy transits from $\omega_{de}>-1$ to $\omega_{de}<-1$ in
agreement with the recent observation which show crossing from
quintessence to phantom phase. Figure $4(b)$ shows that by reduction
of the values of $\eta$, $\omega$ crosses the phantom divide line in
relatively lower values of redshift. In figure $4(c)$ again we
plotted $\omega_{de}$ for $DGP^{(-)}$ branch with $\xi=0.124$\, with
respect to the parameters $s$ and $\eta$.
\newpage

\subsection{Phantom field}

Now we investigate dynamics of a phantom field non-minimally coupled
to induced gravity on the warped DGP brane. Most of the techniques
and discussions for this case are similar to the previous
subsection. The action of the model is
\begin{equation}
{\cal{S}}_{\sigma}=\int_{brane}d^{4}x\sqrt{-g}\Big[-\frac{1}{2}\xi
R\sigma^{2}+\frac{1}{2}\partial_{\mu}\sigma\partial^{\mu}\sigma-V(\sigma)\Big],
\end{equation}
where $\sigma$ is a phantom field. Variation of the action with
respect to $\sigma$ gives the equation of motion of the phantom
field
\begin{equation}
\ddot{\sigma}+3H\dot{\sigma}-\xi R\sigma -\frac{dV}{d\sigma}=0.
\end{equation}
The energy density and pressure of this phantom field are given as
\begin{equation}
\rho_{\sigma}=-\frac{1}{2}\dot{\sigma}^{2}+V(\sigma)+6\xi
H\sigma\dot{\sigma}+3\xi H^{2}\sigma^{2},
\end{equation}
and
\begin{equation}
p_{\sigma}=-\frac{1}{2}\dot{\sigma}^{2}-V(\sigma)-2\xi(\sigma\ddot{\sigma}+2\sigma
H\dot{\sigma}+\dot{\sigma}^{2})-\xi\sigma^2(2\dot{H}+3H^2).
\end{equation}
To compare with the results corresponding to the quintessence field,
we assume the same type of potential
\begin{equation}
V(\sigma)=V_0 \exp(-\lambda\frac{\sigma}{\mu})
\end{equation}
where $V_0$, $\lambda$ and $\mu$ are constant. Differentiation of
the effective energy density of phantom field with respect to
$ln(1+z)$ is given by
$$\frac{d\rho_{de}}{d\ln{(1+z)}}=\frac{3}{\rho_{de}}\Bigg[-\dot{\sigma}^2-2\xi\Big(-H\sigma\dot{\sigma}
+\dot{H}\sigma^2+\sigma\ddot{\sigma}+\dot{\sigma}^2\Big)
+\Big[\dot{\sigma}^2-2\xi\Big(-H\sigma\dot{\sigma}
+\dot{H}\sigma^2+\sigma\ddot{\sigma}+\dot{\sigma}^2\Big)+\rho_{dm}\Big]$$
\begin{equation}
\Big[\varepsilon\eta\Big(A_{0}^2+2\eta\frac{\frac{1}{2}\dot{\sigma}^{2}+V(\sigma)+6\xi
H\sigma\dot{\sigma}+3\xi
H^{2}\sigma^{2}+\rho_{dm}}{\rho_0}\Big)^{-\frac{1}{2}}\Big]\Bigg],
\end{equation}
where $\ddot{\sigma}$ can be deduced from equation of motion of
$\sigma$, (26). On the other hand, Friedmann equation now takes the
following form
$$ (\mu^2+g)^{2}H^4+2f(3\mu^2+g)H^3+\Big[f^2-2l(3\mu^2+g)+2\eta\rho_0 g\Big]H^2$$
\begin{equation}
+\Big(-2fl+\rho_0\eta f\Big)H-2\eta\rho_0(l-\rho_0)-\rho_0^2
A_{0}^2+l^2=0
\end{equation}
where by definition
$$g=-3\xi H^{2}\sigma^{2},$$
$$l=-\frac{1}{2} \dot{\sigma^2}+V(\sigma)+\rho_{dm}+\rho_{0},$$
and $$ f=-6\xi H\sigma\dot{\sigma}.$$ Similar to the last
subsection, there is a forth order equation for $H$ and in principle
this equation has four roots. Two of these roots are un-physical but
two remaining solutions are physical and corresponding to two
branches of solutions in this DGP-inspired model. In table $2$, we
have obtained some acceptable ranges of non-minimal coupling to have
crossing of the phantom divide line in this setup and constraint by
the age of the universe ( that is, we have assumed that the age of
the universe is $13Gyr$).\\

\begin{table}
\begin{center}
\caption{Acceptable range of $\xi$ to have crossing of the phantom
divide line with just one phantom field ( constraint by the age of
the universe). } \vspace{0.5 cm}
\begin{tabular}{|c|c|c|c|c|c|c|c|}
  \hline
  \hline $\varepsilon$&$\xi$ &Acceptable range of $\xi$ & The value of $\xi$ for z=0.25  \\
  \hline +1&negative & $-0.485<\xi\leq0$ & -0.366  \\
  \hline +1&positive & $0.055\leq\xi\leq 0.170$ & 0.088 \\
 \hline -1&negative &no crossing &--- \\
  \hline-1&positive &$0\leq\xi<0.166$&0.166  \\
 \hline
\end{tabular}
\end{center}
\end{table}
Figure $5(a)$ shows that for self-accelerating branch of the model (
with $\varepsilon=+1$) and with negative values of the non-minimal
coupling, there is a crossing of the phantom divide line by the
equation of state parameter. For $\xi=-0.366$, we have a crossing at
the point with $s=-0.22$ corresponding to $z=0.25$ in agreement with
observations. Here the EoS runs from above $-1$ to below $-1$ ( from
quintessence to phantom phase). Figure $5(b)$ shows that for
sufficiently small values of $\eta$, equation of state parameter,
$\omega$, crosses the phantom divide line in relatively large values
of redshift. In figure $5(c)$ we plotted $\omega_{de}$ for
$DGP^{(+)}$ branch with $\xi=-0.366$ and with respect to the
parameters $s$ and $\eta$.\\

\begin{figure}[htp]
\begin{center}
\includegraphics{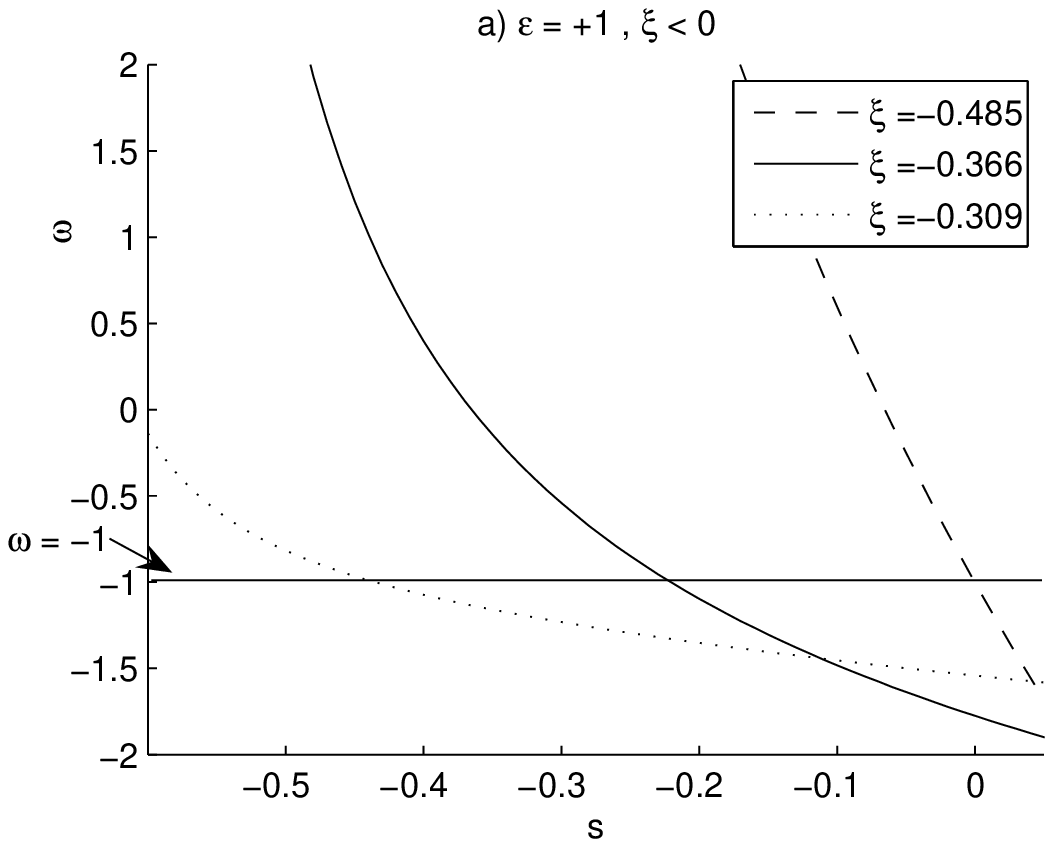} \vspace{5cm}\includegraphics{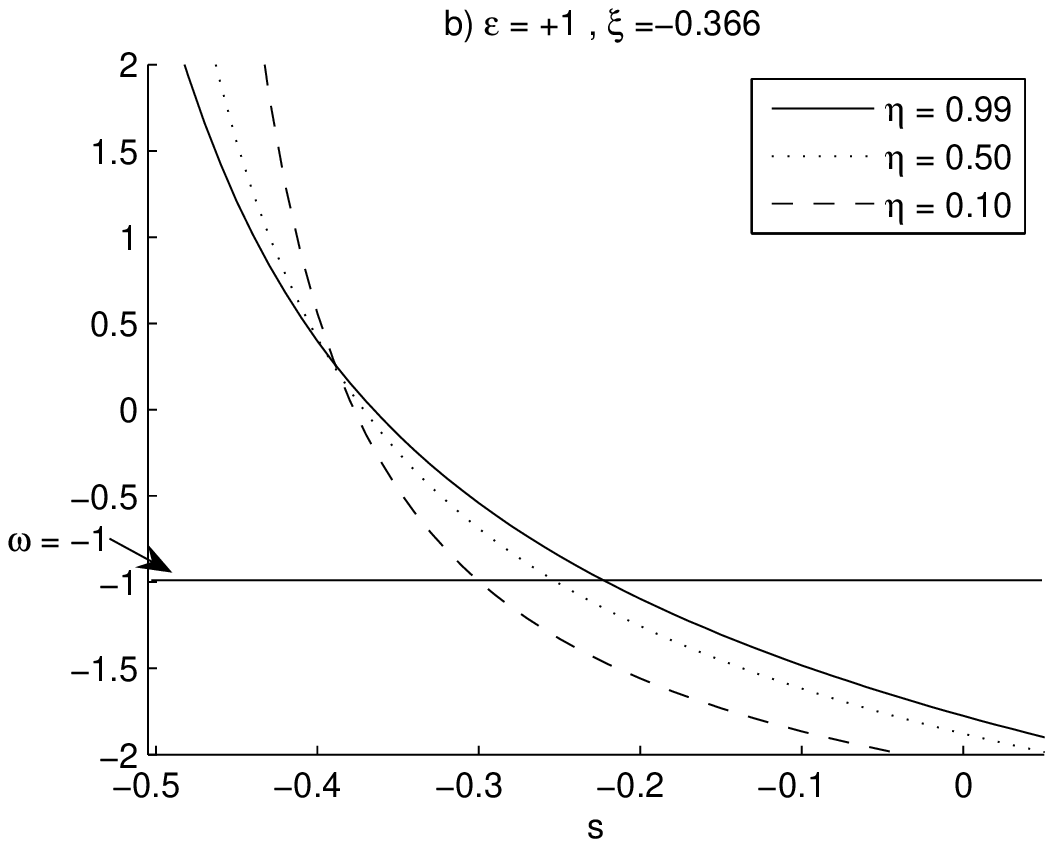}\vspace{5cm}\includegraphics{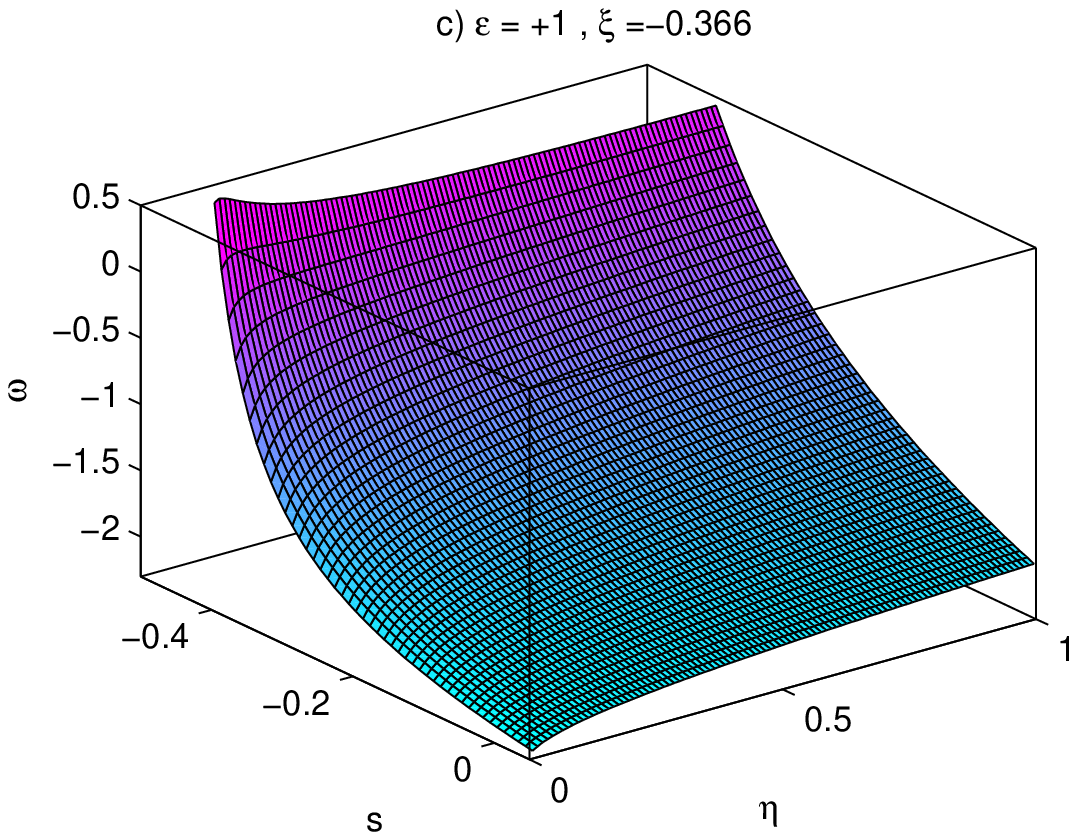}
\end{center}
\vspace{2.5cm}
 \caption{\small {Dynamics of the equation of state parameter with a phantom
 field on the self-accelerating branch of the model. a) With negative values of the nonminimal
 coupling, the EoS of dark energy crosses $\omega=-1$ line running from quintessence to phantom phase.
 For instance, with $\xi=-0.366$, this crossing occurs at $s=-0.22$ or $z=0.25$.
 b)The role played by $\eta$ in equation of state of phantom field on the self-accelerating
 branch of the model. For sufficiently small values of $\eta$, equation of state
parameter crosses the phantom divide line in relatively large values
of redshift ( in contrast to the case with quintessence field). For
example, with $\xi=-0.366$, the EoS of dark energy crosses
$\omega=-1$ line with $\eta=0.99$ at $s\approx-0.22$ or $z=0.25$,
while for $\eta=0.50$ this crossing occurs at $s\approx-0.250$ or
$z\approx0.280$ and for $\eta=0.10$ this occurs at $s\approx-0.297$
or $z=0.345$. c) Equation of state parameter, $\omega$, versus $s$
and $\eta$ with $\xi=-0.366$ in self-accelerating branch of the
model in a 3-dimensional plot. In the self-accelerating branch with
phantom field, $\omega=-1$ line crossing runs from quintessence to
phantom phase. }}
\end{figure}

Figure $6(a)$ shows that for self-accelerating branch of this model
and with positive values of $\xi$, there is a crossing of the
phantom divide line too. For example, with $\xi=0.088$, we have a
crossing at the point with $s=-0.22$ corresponding to $z=0.25$ in
agreement with observational data. As another important point, the
EoS parameter transits from $\omega_{de}>-1$ to $\omega_{de}<-1$.
Figure $6(b)$ shows that by decreasing the values of $\eta$,
equation of state parameter, $\omega$, crosses the phantom divide
line in relatively large values of redshift. In figure $6(c)$ we
plotted $\omega_{de}$ for $DGP^{(+)}$ branch with $\xi=0.088$ and
with respect to the parameters $s$ and $\eta$.
\begin{figure}[htp]
\begin{center}
\includegraphics{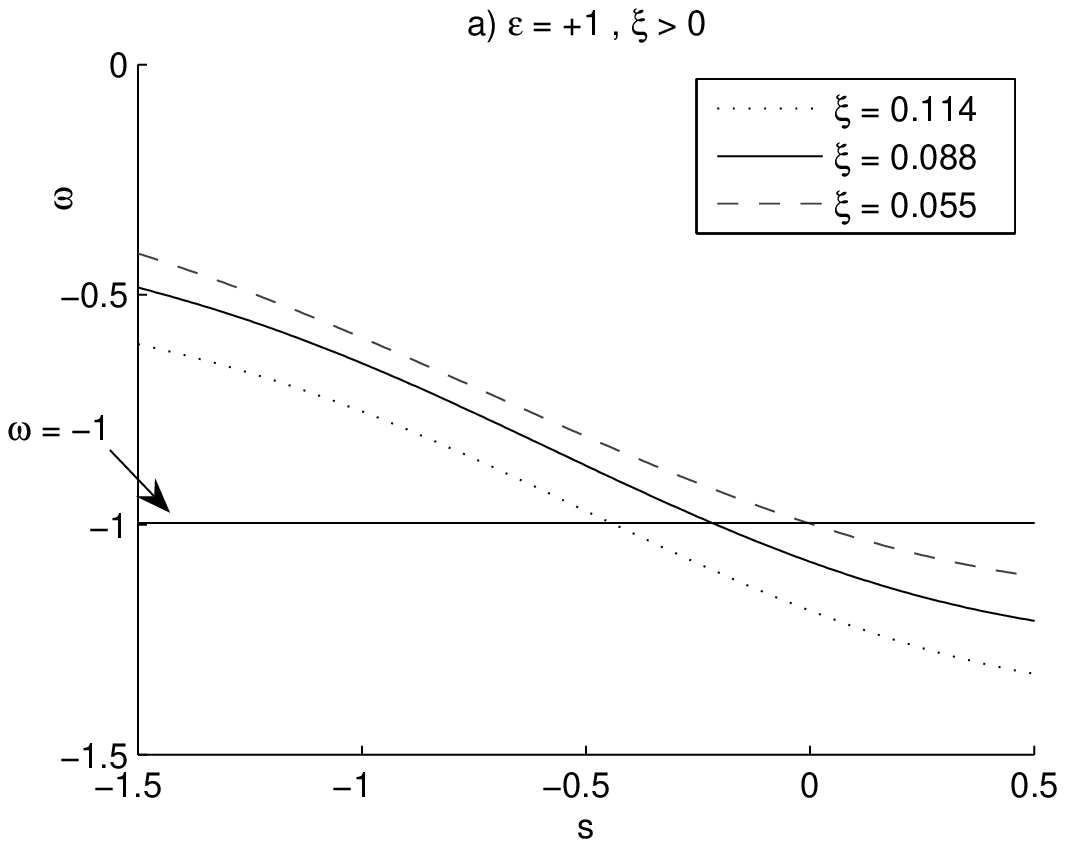} \vspace{5cm}\includegraphics{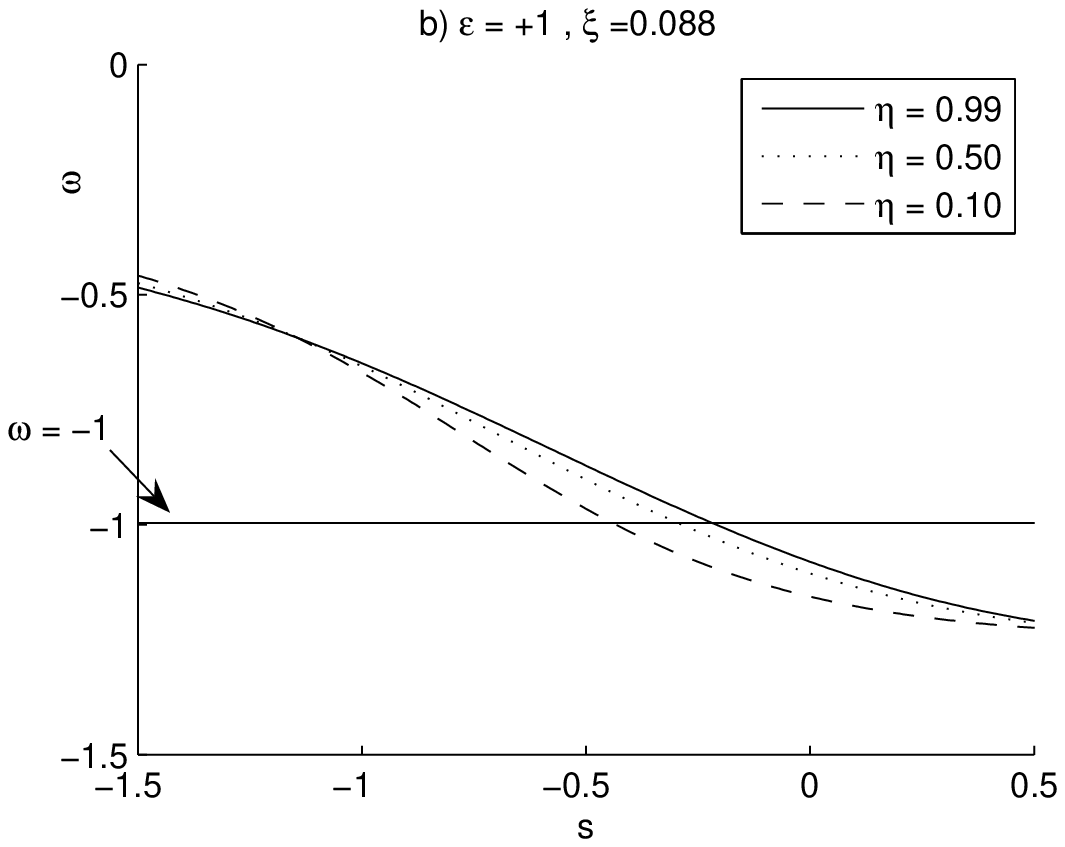}\vspace{5cm}\includegraphics{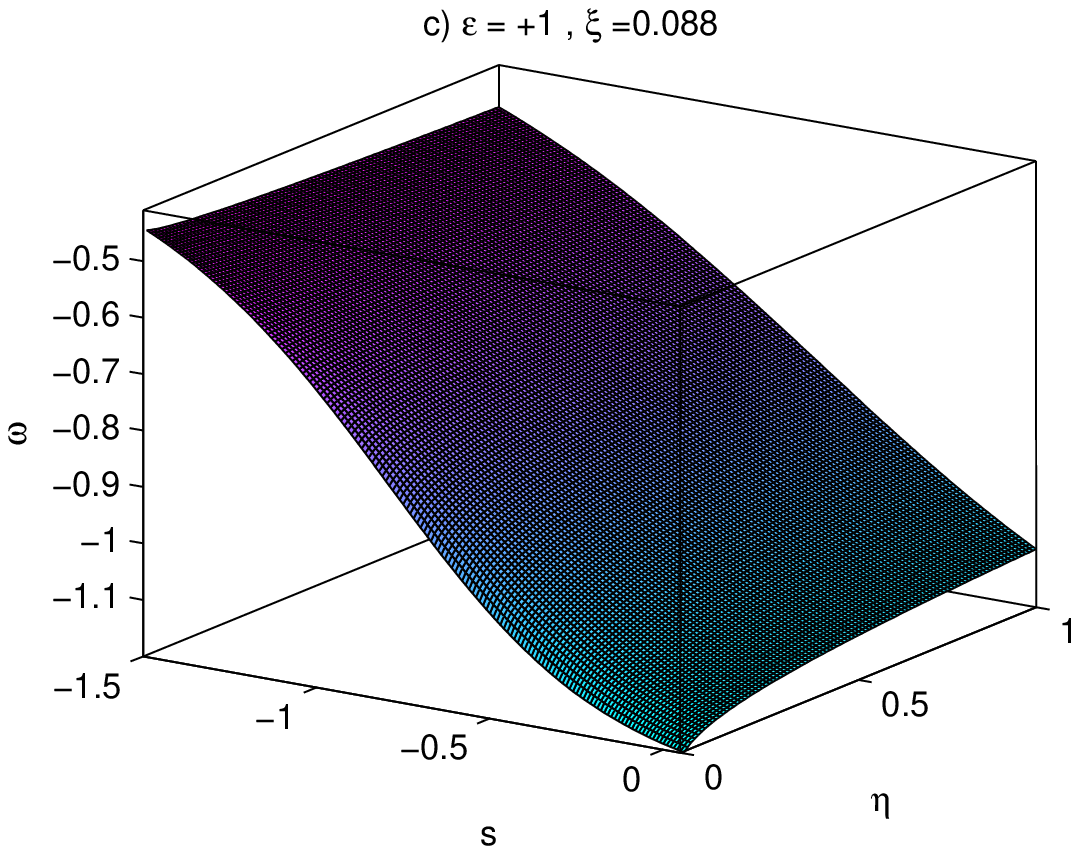}
\end{center}
\vspace{2.5cm}
 \caption{\small {The self-accelerating branch of the model with positive values
 of the non-minimal coupling in the presence of just one phantom field on the brane.
 The situation is similar to previous figure and phantom divide line crossing
 runs from quintessence to phantom phase.  a) With positive values of the nonminimal
 coupling, the EoS of dark energy crosses $\omega=-1$ line for
 $\xi=0.088$ at $s=-0.22$ or $z=0.25$. b)The role of $\eta$ on the
 crossing of the phantom divide line. As previous figure and contrary to
 quintessence case, for sufficiently small values of $\eta$ equation of state parameter
crosses the phantom divide line in relatively large values of
redshift. For example, with $\xi=0.088$, the EoS of dark energy
crosses $\omega=-1$ line with $\eta=0.99$ at $s\approx-0.22$ or
$z=0.25 $, while for $\eta=0.5$ this crossing occurs at
$s\approx-0.29$ or $z=0.33$ and for $\eta=0.1$ this occurs at
$s\approx-0.44$ or $z=0.55$. c) A 3-dimensional plot of the equation
of state parameter, $\omega$, versus $s$ and $\eta$ with
$\xi=0.088$. }}
\end{figure}
In figure $7$ we see that for normal or non self-accelerating branch
of the model and with negative values of the non-minimal coupling,
there is no crossing of the phantom divide line.\\
\begin{figure}[htp]
\begin{center}
\includegraphics{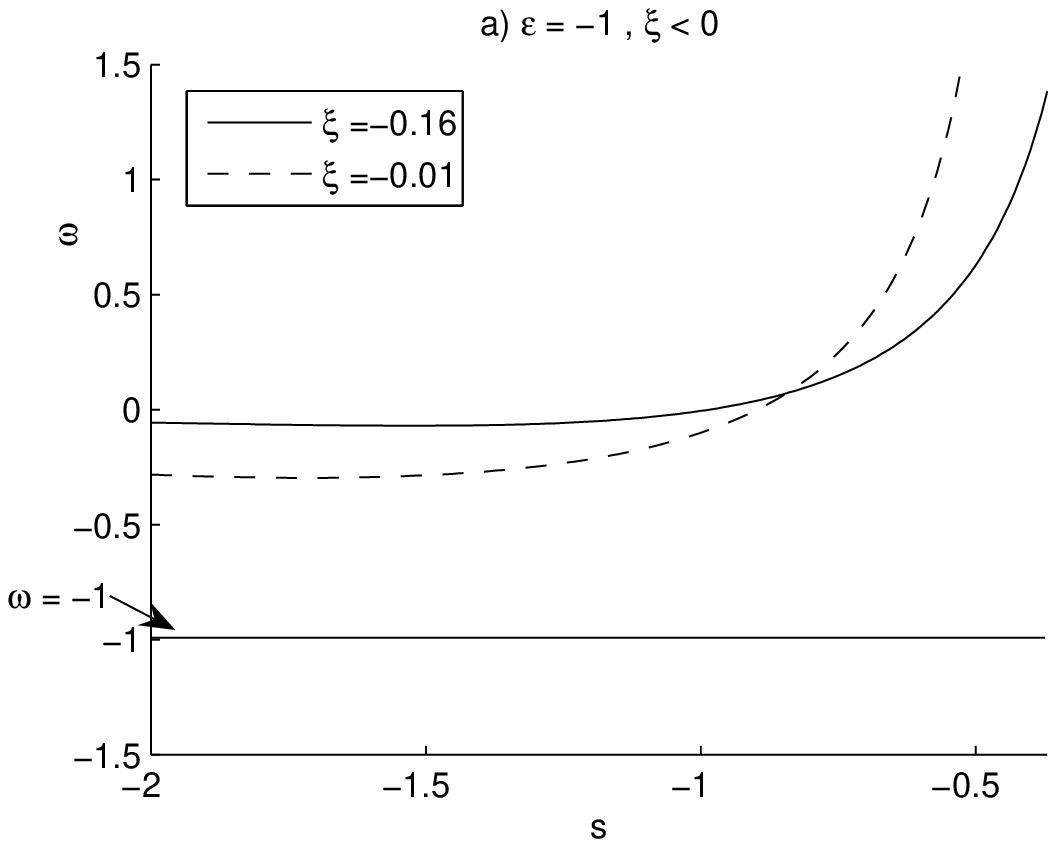}
\end{center}
\vspace{14 cm}
 \caption{\small {With a phantom field on the warped DGP brane, there is no crossing of the phantom
 divide line in normal ( non self-accelerating ) branch of the model with negative values of
 the non-minimal coupling. Comparison with figure $2$ for a quintessence
 field on the warped DGP brane, shows the differences between two situation.}}
\end{figure}
Figure $8(a)$ shows the phantom divide line crossing of EoS
parameter for normal branch ($\varepsilon=-1$) of the model with
positive values of $\xi$. Here, the EoS of dark energy transits from
$\omega_{de}>-1$ to $\omega_{de}<-1$. Figure $8(b)$ shows that by
reduction of the values of $\eta$, $\omega$ crosses the phantom
divide line in relatively smaller values of redshift. In figure
$8(c)$ again, we plotted $\omega_{de}$ for normal branch of the
model with $\xi=0.166$ versus the parameters $s$ and $\eta$.

\begin{figure}[htp]
\begin{center}
\includegraphics{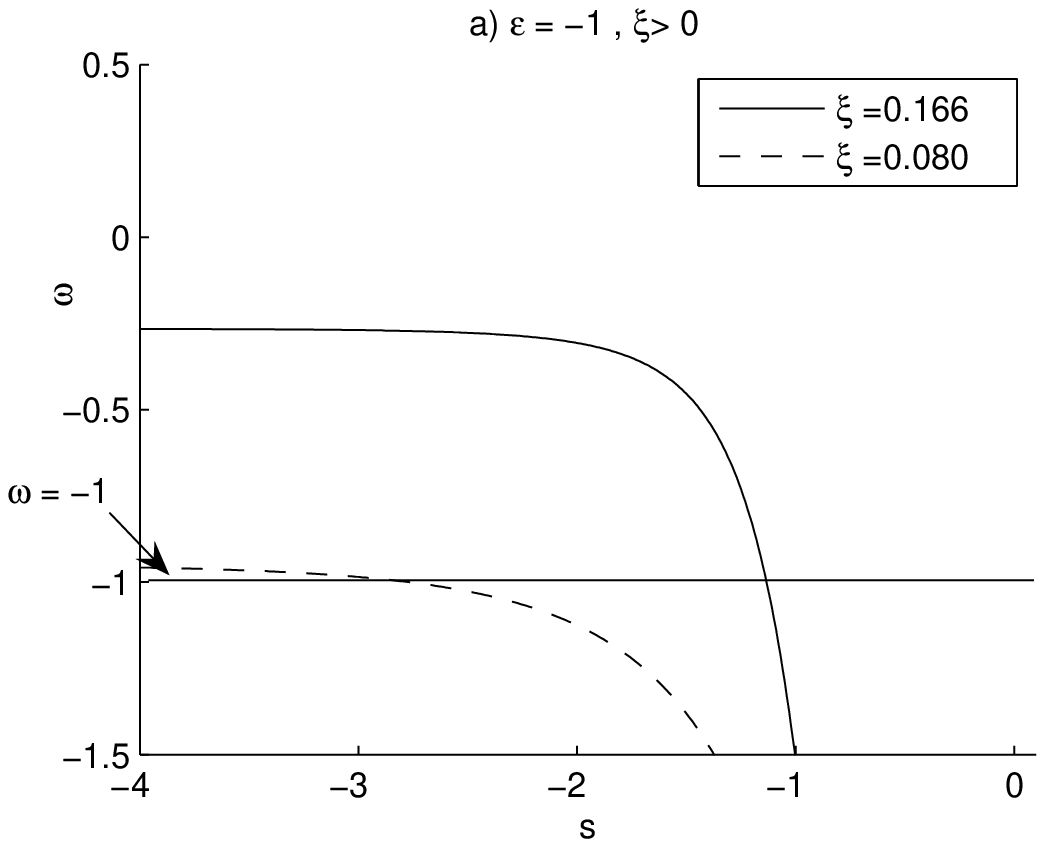} \vspace{5cm}\includegraphics{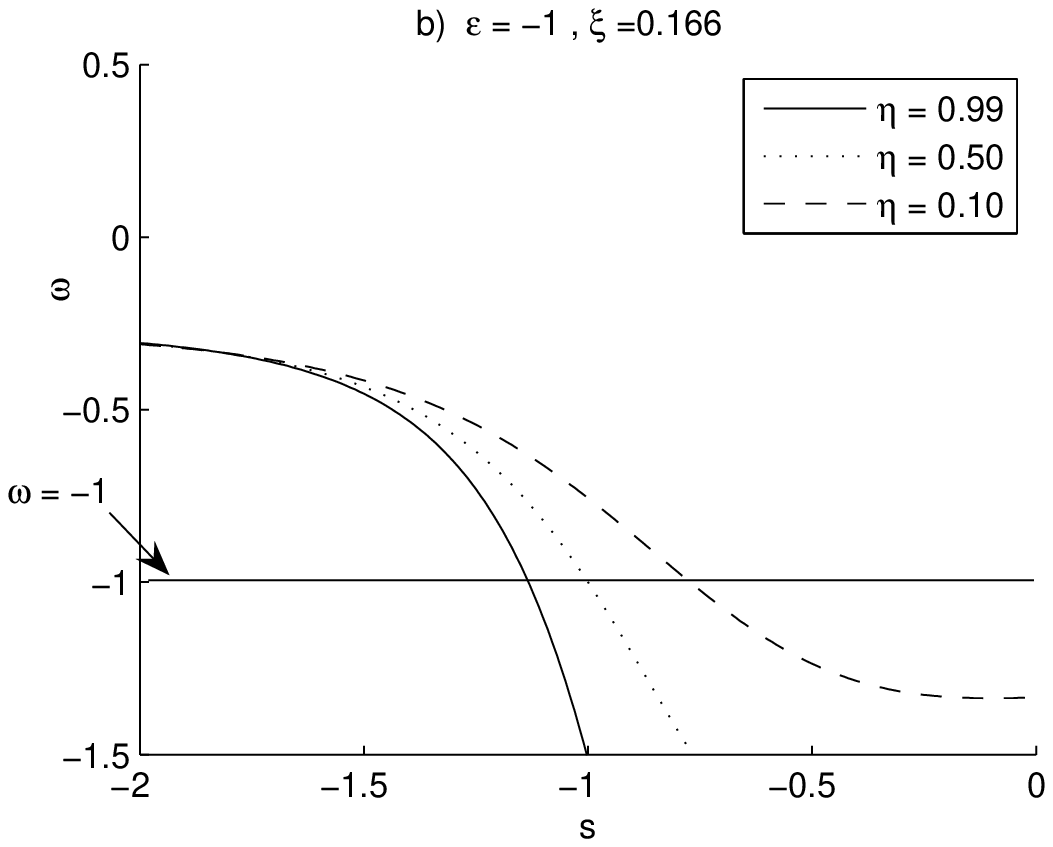}\vspace{5cm}\includegraphics{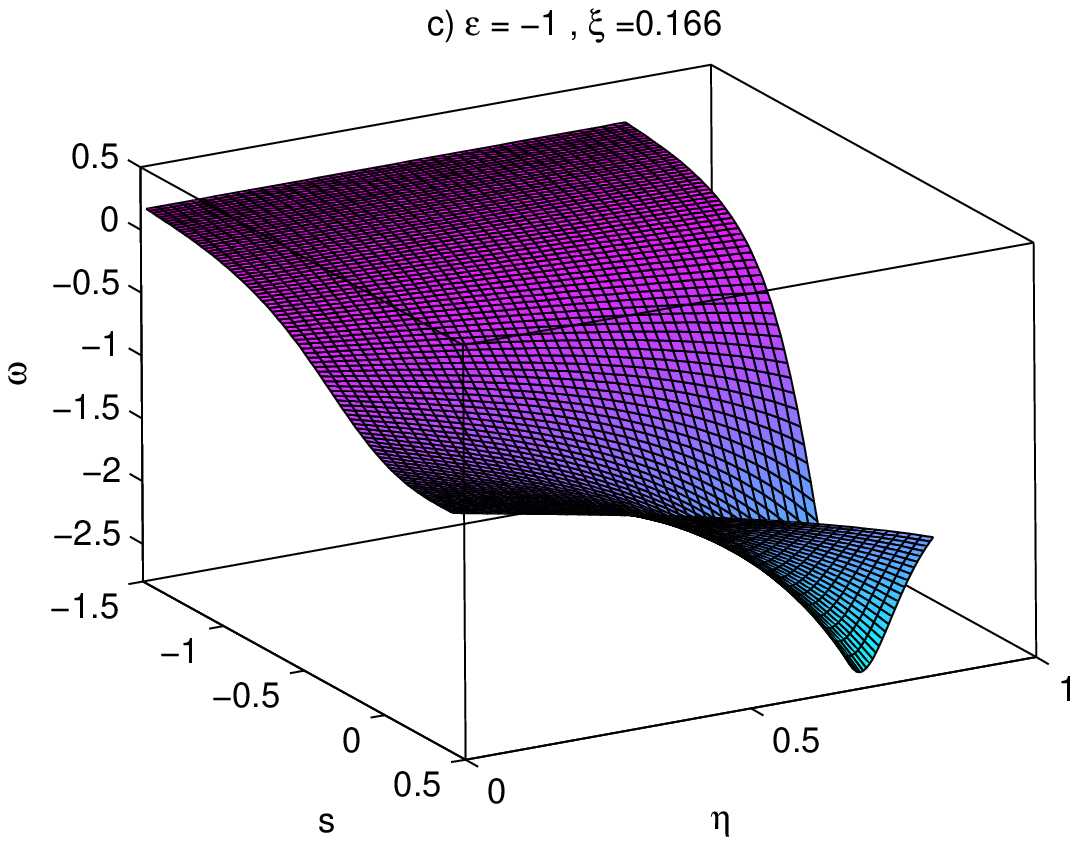}
\end{center}
\vspace{2.5cm}
 \caption{\small {In the normal branch of the model
 with a phantom field on the brane, crossing of the phantom divide line runs
 from quintessence to phantom phase. This is a general behavior for equation of
 state of phantom field on the brane independent
 of the signs of non-minimal coupling and $\varepsilon$. This is supported by observations too. However,
 as we have shown in figures $1$, $3$ and $4$, the situation for
 quintessence field on the warped DGP brane is different and crossing of the phantom divide
 line depends on the signs of the non-minimal coupling and $\varepsilon$.
 That is, while with phantom field on the warped DGP brane,
 crossing of the phantom divide line runs always from quintessence to phantom phase,
 in the case of quintessence field on the brane this running of crossing depends on the
 sign of the non-minimal coupling and $\varepsilon$. }}
\end{figure}

\newpage
\section{Summary and Conclusion}
An alternative approach to explain current positively accelerated
phase of the universe expansion is to use a multi-component dark
energy with at least one non-canonical phantom field. The analysis
of the properties of dark energy from recent observations mildly
favor models where $\omega=\frac{p}{\rho}$ crosses the phantom
divide line, $\omega=-1$, in the near past. In this paper, we have
considered a scalar field non-minimally coupled to induced gravity
on the warped DGP braneworld as a dark energy component and we have
investigated the roles played by the non-minimal coupling and the
warp effect on the dynamics of the equation of state parameter. In
this respect, we have studied the dynamics of equation of state
parameter focusing on the crossing of the phantom divide line in
this setup. As it is well-known, in the absence of scalar field
there is no crossing of the phantom divide line in self-accelerating
branch of the DGP setup and this is the case even in the warped DGP
scenario. However, in the presence of a scalar field ( minimally or
non-minimally coupled to induced gravity), it is possible to realize
this crossing. We have shown that this crossing is possible for a
suitable range of the model parameters and especially for some
specific values of the non-minimal coupling and parameter $\eta$
related to the warp effect in this DGP-inspired scenario. In the
first stage, we have considered a canonical (quintessence) scalar
field non-minimally coupled to the induced gravity on the warped DGP
brane. In this case, we have shown that crossing of the cosmological
constant line by the EoS parameter of the quintessence field occurs
in both self-accelerating and normal branches of the model. For
self-accelerating branch of the model( with $\varepsilon=+1$ ),
crossing of the cosmological constant line occurs with negative
values of the non-minimal coupling parameter and this crossing runs
from phantom to quintessence phase. There is no crossing behavior in
the self-accelerating branch of the model with positive values of
the non-minimal coupling. On the other hand, for normal branch of
the model( with $\varepsilon=-1$ ), the equation of state parameter
of dark energy crosses the phantom divide line with negative values
of the non-minimal coupling as well as its positive values, but the
crossing behavior is completely different from the former one ( the
self-accelerating branch ). Indeed, in this case the EoS parameter
of dark energy crosses the phantom divide line in a different
direction; from quintessence to phantom phase and this is supported
by recent observations. By investigating the role played by the
parameter $\eta$ ( which is related to the warp effect) in both
branches of the model, we found that decreasing of the effect of
$\eta$ factor leads to the result that the EoS parameter of dark
energy crosses the cosmological constant line in relativity smaller
values of redshift.  We should stress here that negative values of
the non-minimal coupling are interesting at least theoretically
since they show anti-gravitation. However, recent observational
constraints on the values of the non-minimal coupling favor
positivity of this factor ( see for instance [28] and [29]).

In the next stage, we have considered a phantom field non-minimally
coupled to the induced gravity on the warped DGP brane. We have
shown that in the self-accelerating branch of the model, crossing of
the phantom divide line by the EoS parameter of dark energy occurs
with both signs of the non-minimal coupling and this crossing runs
always from quintessence to phantom phase. Finally, we have shown
that with a phantom field on the warped DGP brane, there is no
crossing of the phantom divide line in the normal ( non
self-accelerating ) branch of the model with negative values of the
non-minimal coupling.  In the normal branch with phantom field, by
considering positive values of the non-minimal coupling parameter,
the EoS parameter of dark energy crosses the phantom divide line
from quintessence to phantom phase supported by observations. With a
phantom field on the self-accelerating branch of the model and for
both signs of the non-minimal coupling, reduction of the values of
$\eta$ leads to phantom divide line crossing in relatively larger
values of redshift, but in normal branch reduction of $\eta$ leads
to crossing in smaller values of redshift.

In summary, with positive values of the non-minimal coupling which
is physically more relevant, we have shown that: in the
self-accelerating branch of this warped-DGP setup with just one
quintessence field, crossing the phantom divide line cannot be
realized. However, it is possible to realize phantom divide line
crossing in the normal branch of the model and this crossing runs
from phantom to quintessence phase. With just one phantom field on
the brane, it is possible to realize phantom divide line crossing
with positive values of the non-minimal coupling in both branches of
this DGP-inspired model and this crossing runs from quintessence to
phantom phase. Although with a phantom field on the warped DGP
brane, crossing of the phantom divide line runs always from
quintessence to phantom phase, in the case of quintessence field on
the brane this running depends on the sign of the non-minimal
coupling and $\varepsilon$. Finally, we should stress that
self-accelerating branch of the DGP scenario suffers from ghost
instabilities( see [30,31]). Incorporation of new degrees of freedom
such as the non-minimal coupling of the scalar field and induced
gravity and also warp geometry of the bulk provides a wider
parameters space in our setup and this wider parameter space may
provide a suitable basis to treat ghost instabilities.

{\bf Acknowledgment}\\
This work has been supported partially by Research Institute for
Astronomy and Astrophysics of Maragha, IRAN.

\end{document}